\documentclass[prl,twocolumn,floatix,amssymb,showpacs,amsmath,superscriptaddress]{revtex4-1}
\usepackage{graphicx}
\usepackage{epsfig,color}
\usepackage{amsmath,amssymb}
\begin{document}

\newcommand{\ket}[1]{\ensuremath{\left|{#1}\right\rangle}}
\newcommand{\bra}[1]{\ensuremath{\left\langle{#1}\right|}}
\newcommand{\quadr}[1]{\ensuremath{{\not}{#1}}}
\newcommand{\quadrd}[0]{\ensuremath{{\not}{\partial}}}
\newcommand{\slpar}{\partial\!\!\!/}
\newcommand{\gtrescero}{\gamma_{(3)}^0}
\newcommand{\gtresuno}{\gamma_{(3)}^1}
\newcommand{\gtresi}{\gamma_{(3)}^i}

\title{Embedding Quantum Simulators for Quantum Computation of Entanglement}

\author{R. Di Candia} 
\affiliation{Department of Physical Chemistry, University of the Basque Country UPV/EHU, Apartado 644, 48080 Bilbao, Spain}
\author{B. Mejia}
\affiliation{Departamento de Ciencias, Pontificia Universidad Cat\'olica del Per\'u, Apartado 1761, Lima, Per\'u}
\author{H. Castillo}
\affiliation{Departamento de Ciencias, Pontificia Universidad Cat\'olica del Per\'u, Apartado 1761, Lima, Per\'u}
\author{J. S. Pedernales}
\affiliation{Department of Physical Chemistry, University of the Basque Country UPV/EHU, Apartado 644, 48080 Bilbao, Spain}
\author{J. Casanova}
\affiliation{Department of Physical Chemistry, University of the Basque Country UPV/EHU, Apartado 644, 48080 Bilbao, Spain}
\author{E. Solano}
\affiliation{Department of Physical Chemistry, University of the Basque Country UPV/EHU, Apartado 644, 48080 Bilbao, Spain}
\affiliation{IKERBASQUE, Basque Foundation for Science, Alameda Urquijo 36, 48011 Bilbao, Spain}

\date{\today}
   
\begin{abstract}
We introduce the concept of embedding quantum simulators, a paradigm allowing the efficient quantum computation of a class of bipartite and multipartite entanglement monotones. It consists in the suitable encoding of a simulated quantum dynamics in the enlarged Hilbert space of an embedding quantum simulator. In this manner, entanglement monotones are conveniently mapped onto physical observables, overcoming the necessity of full tomography and reducing drastically the experimental requirements. Furthermore, this method is directly applicable to pure states and, assisted by classical algorithms, to the mixed-state case. Finally, we expect that the proposed embedding framework paves the way for a general theory of enhanced one-to-one quantum simulators. 
\end{abstract}


\maketitle

Entanglement is considered  one of the most remarkable features of quantum mechanics~\cite{Nielsen, Horodecki1}, stemming from bipartite or multipartite correlations without classical counterpart. Firstly revealed by Einstein, Podolsky, and Rosen as a possible drawback of quantum theory~\cite{Einstein}, entanglement was subsequently identified as a fundamental resource for quantum communication~\cite{Ekert91, Bennett95} and quantum computing purposes~\cite{Shor95, Gottesman99}. Beyond considering entanglement  as a purely theoretical feature, the development of quantum technologies has allowed us to create, manipulate, and detect entangled states in different quantum platforms. Among them, we can mention trapped ions, where eight-qubit W and fourteen-qubit GHZ states have been created~\cite{Haffner05, Monz11}, circuit QED (cQED) where  seven superconducting elements have been entangled~\cite{Mariantoni11}, superconducting circuits where continuous-variable entanglement has been realized in propagating quantum microwaves~\cite{Menzel12}, and bulk-optic based setups  where entanglement between eight photons has been achieved~\cite{Gao10}.

\begin{figure}[t]
\begin{center}
\vspace{0.5cm}
\hspace{-0.3cm}
\includegraphics [width= 0.97 \columnwidth]{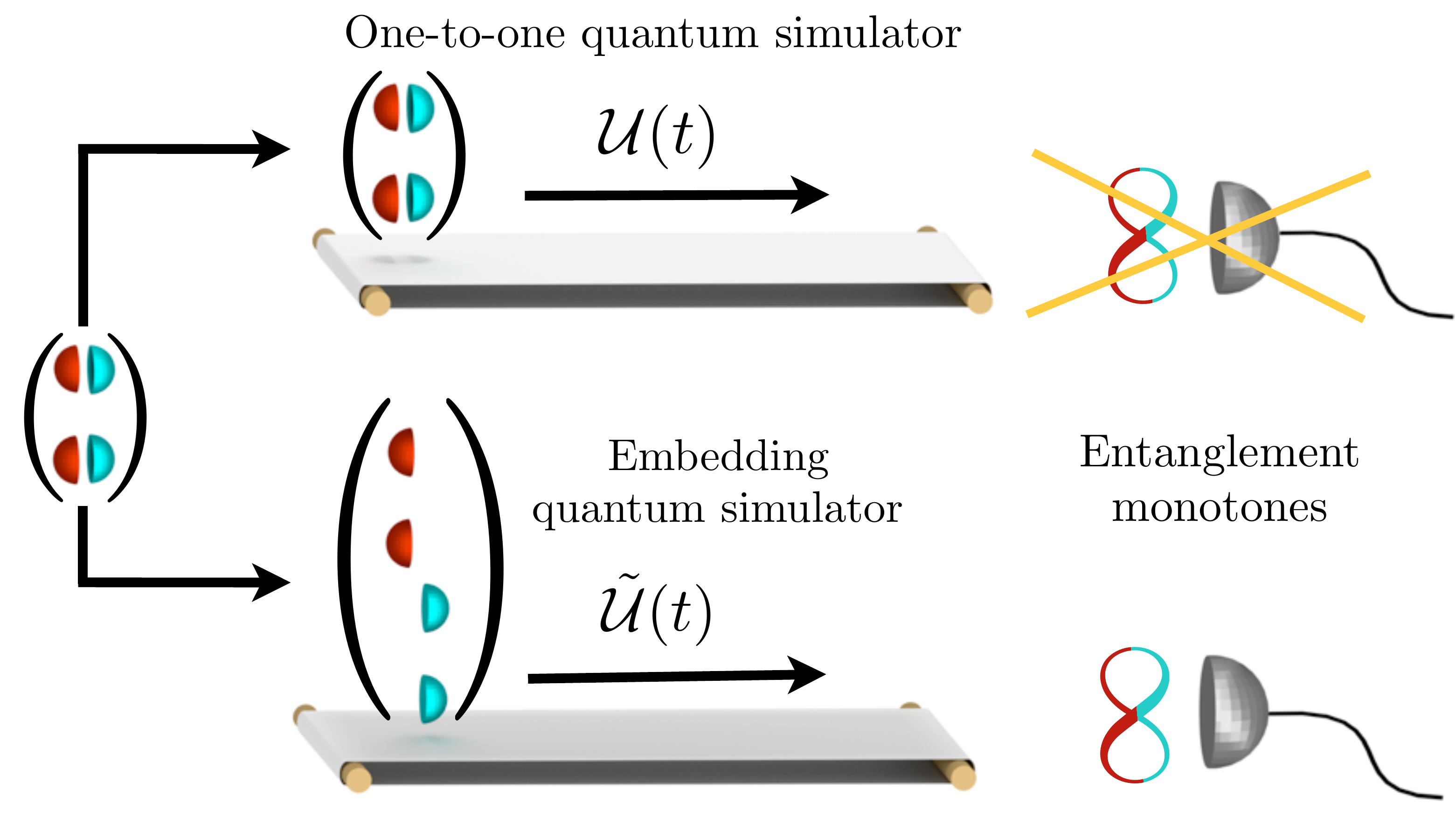}
\end{center}
\caption{(color online) One-to-one quantum simulator versus embedding quantum simulator. The conveyor belts represent the dynamical evolution of the quantum simulators. The real (red) and imaginary (blue) parts of the simulated wave vector components are split in the embedding quantum simulator, allowing the efficient computation of entanglement monotones.}\label{figembedding}
\end{figure}

Quantifying entanglement is considered a particularly difficult task, both from theoretical and experimental viewpoints. In fact, it is challenging to define entanglement measures for an arbitrary number of parties~\cite{Wong01, Barnum04}. Moreover, the existing entanglement monotones~\cite{Vidal00} do not correspond directly to the expectation value of a Hermitian operator~\cite{Wootters98}. Accordingly,  the computation of many entanglement measures, see Ref.~\cite{Guhne1} for lower bound estimations, requires  previously the reconstruction of the full quantum state, which could be a cumbersome problem if the size of the associated Hilbert space is large. If we consider, for instance, an $N$-qubit system, quantum tomography techniques become already experimentally unfeasible for $N\sim10$ qubits. This is because the dimension of the Hilbert space grows exponentially with $N$, and the number of observables needed to reconstruct the quantum state scales~as~$2^{2N}-1$. 

From a general point of view, a {\it standard quantum simulation} is meant to be implemented in a {\it one-to-one quantum simulator} where, for example, a two-level system in the simulated dynamics is directly represented by another two-level system in the simulator. In this Letter, we introduce the concept of {\it embedding quantum simulators}, allowing the efficient computation of a wide class of entanglement monotones~\cite{Vidal00}. This method can be applied at any time of the evolution of a simulated bipartite or multipartite system, with the prior knowledge of the Hamiltonian $H$ and the corresponding initial state $|\psi_0\rangle$. The efficiency of the protocol lies in the fact that, unlike standard quantum simulations, the evolution of  the state $|\psi_0\rangle$ is embedded in an {\it enlarged Hilbert space} dynamics (see Fig.~\ref{figembedding}). Note that enlarged-space structures have been previously considered for different purposes in Refs.~\cite{Rudolph1, Fernandez1, McKague1, Alvarez}. In our case, antilinear operators associated with a certain class of entanglement monotones can be efficiently encoded into physical observables, overcoming the necessity of full state reconstruction. The simulating quantum dynamics, which embeds the desired quantum simulation, may be implemented in different quantum technologies with analog and digital simulation methods.

{\it Complex conjugation and entanglement monotones.---}
An entanglement monotone is a function of the quantum state, which is zero for all separable states and does not increase on average under local quantum operations and classical communication ~\cite{Vidal00}. There are several functions satisfying these basic properties, as concurrence~\cite{Wootters98} or three-tangle~\cite{Dur00}, extracting information about a specific feature of entanglement. For pure states, an entanglement monotone $E(|\psi\rangle)$ can be defined univocally, while the standard approach for mixed states requires the computation of the convex roof 
\begin{equation}\label{roof}
E(\rho)=\min_{\{p_i,|\psi_i\rangle\}}\;\sum_{i}p_iE(|\psi_i\rangle).
\end{equation}
Here, $\rho=\sum_{i}p_i|\psi_i\rangle\langle\psi_i|$ is the density matrix describing the system, and the minimum in Eq.~(\ref{roof}) is taken over all possible pure-state decompositions~\cite{Horodecki1}. 

A systematic procedure to define entanglement monotones for pure states involves  the complex-conjugation operator $K$~\cite{Osterloh1, Uhlmann07}. For instance, the concurrence for  two-qubit pure states~\cite{Wootters98} can be written as
\begin{equation}
\label{Concurrence}
C(|\psi\rangle)\equiv|\langle\psi|\sigma_y\otimes\sigma_y K|\psi\rangle | .
\end{equation}
Note that $\sigma_y\otimes\sigma_yK$,  where $K | \psi \rangle \equiv | \psi^* \rangle$, is an antilinear operator that cannot be associated with a physical observable. In general, we can construct entanglement monotones for  $N$-qubit systems  combining three operational building blocks: $K$,  $\sigma_y$, and $g^{\mu\nu}\sigma_\mu\sigma_\nu$, with $g^{\mu\nu}=\text{diag}\{-1,1,0,1\}$, $\sigma_0=\mathbb{I}_2$, $\sigma_1=\sigma_x$, $\sigma_2=\sigma_y$, $\sigma_3=\sigma_z$,   where we assume the repeated index summation convention~\cite{Osterloh1}. For a two-qubit system, $N=2$, we can define $|\langle\psi|\sigma_y\otimes\sigma_yK|\psi\rangle|$ and  $|g^{\mu\nu}g^{\lambda\tau}\langle\psi|\sigma_\mu\otimes\sigma_\lambda K|\psi\rangle\langle\psi|\sigma_\nu\otimes\sigma_\tau K|\psi\rangle|$  as entanglement monotones. The first expression corresponds to the concurrence and the second one is a second-order monotone defined in Ref.~\cite{Osterloh1}. For $N=3$ we have $|g^{\mu\nu}\langle\psi|\sigma_\mu\otimes\sigma_y\otimes\sigma_yK|\psi\rangle\langle\psi|\sigma_\nu\otimes\sigma_y\otimes\sigma_yK|\psi\rangle|$, corresponding to the $3$-tangle~\cite{Dur00}, and so on. 

To evaluate the above class of entanglement monotones in a one-to-one quantum simulator, we would need to perform full tomography on the system. This is because  terms like $\langle\psi|OK|\psi\rangle\equiv \langle\psi|O|\psi^*\rangle$, with $O$ Hermitian, do not correspond to  the expectation value of a physical observable, and they have to be computed classically once each complex component of $|\psi\rangle$ is known. We will explain now how to compute efficiently quantities as $\langle\psi| O K|\psi\rangle$ in our proposed embedding quantum simulator, via the measurement of a reduced number of observables.

Consider a pure quantum state $|\psi\rangle$ of an $N$-qubit system $\in \mathbb{C}_{2^N}$, whose evolution is governed  by the Hamiltonian $H$ via the Schr\"odinger equation ($\hbar=1$)
\begin{equation}\label{Schro}
(i  \partial_{t} -H )|\psi(t)\rangle=0.
\end{equation}
The quantum dynamics associated with the Hamiltonian $H$ can be implemented in a one-to-one quantum simulator~\cite{Feynman82, Lloyd96} or, alternatively, it can be encoded in an embedding quantum simulator, where $K$  may become a physical quantum operation~\cite{CasanovaX}. The latter can be achieved according to the following rules.

{\it Embedding quantum simulator.---}  We define a mapping $\mathcal{M}:\mathbb{C}_{2^N}\rightarrow\mathbb{R}_{2^{N+1}}$ in the following way:
\begin{equation}\label{map}
|\psi\rangle=\left( \begin{array}{c} \psi_{\rm{re}}^1+i\psi_{\rm{im}}^1\\ \psi_{\rm{re}}^2+i\psi_{\rm{im}}^2\\ \psi_{\rm{re}}^3+i\psi_{\rm{im}}^3\\  \vdots\end{array} \right) {\huge\xrightarrow {\mathcal{M}}} \,\, |\tilde \psi\rangle=\left( \begin{array}{c} \psi_{\rm{re}}^1\\ \psi_{\rm{re}}^2\\ \psi_{\rm{re}}^3\\  \vdots\\ \psi_{\rm{im}}^1 \\  \psi_{\rm{im}}^2 \\  \psi_{\rm{im}}^3 \\  \vdots\end{array} \right).
\end{equation}
Hereafter, we will call $\mathbb{C}_{2^N}$ the {\it simulated space}  and $\mathbb{R}_{2^{N+1}}$ the {\it simulating space} or the {\it enlarged space}. We note that the resulting vector  $|\tilde{\psi}\rangle$ has only real components (see refs.~\cite{Rudolph1, Fernandez1, McKague1} for other developments involving real Hilbert spaces), and that the reverse mapping is  $|\psi\rangle=M|\tilde\psi\rangle$, with $M=\left(1\;,\; i\right)\otimes\mathbb{I}_{2^N}$. It is noteworthy to mention that, for an unknown initial state, the mapping $\mathcal M$ is not physically implementable. However, according to Eq.~(\ref{map}), the knowledge of the initial state in the simulated space determines completely the possibility of initializing the state in the enlarged space. Furthermore, it can be easily checked that the inverse mapping $M$ can always be completed to form a unitary operation.

Now, we can write
\begin{equation}\label{relation}
K|\psi\rangle \equiv|\psi^*\rangle=M|\tilde\psi^*\rangle= M(\sigma_z\otimes\mathbb{I}_{2^N})|\tilde\psi\rangle\equiv M\tilde K|\tilde\psi\rangle ,
\end{equation}
which, despite its simple aspect, has important consequences. Basically, Eq.~(\ref{relation})  tells us that while $|\psi\rangle$ and $|\psi^*\rangle$ are connected by the unphysical operation $K$ in the simulated space, the relation between their images in the enlarged space, $|\tilde{\psi}\rangle$ and $|\tilde{\psi}^*\rangle$, is a physical quantum gate $\tilde{K}\equiv(\sigma_z\otimes\mathbb{I}_{2^N})$. In this way, we obtain that
\begin{equation}\label{obs}
\langle\psi|OK|\psi\rangle=\langle\tilde\psi|M^\dag O M (\sigma_z\otimes\mathbb{I}_{2^N})|\tilde \psi\rangle,
\end{equation}
where we can prove that
\begin{equation}
M^\dag  O M (\sigma_z\otimes\mathbb{I}_{2^N}) = (\sigma_z-i\sigma_x)\otimes O.
\end{equation}
Note that $M^\dag  O M (\sigma_z\otimes\mathbb{I}_{2^N})$ is a linear combination of Hermitian operators $\sigma_z\otimes O$ and $\sigma_x\otimes O$. Hence, its expectation  value can be efficiently computed via the measurement  of these two observables in the enlarged space.

\begin{figure}[t]
\begin{center}
\vspace{0.5cm}
\hspace{-0.3cm}
\includegraphics [width= 1.01 \columnwidth]{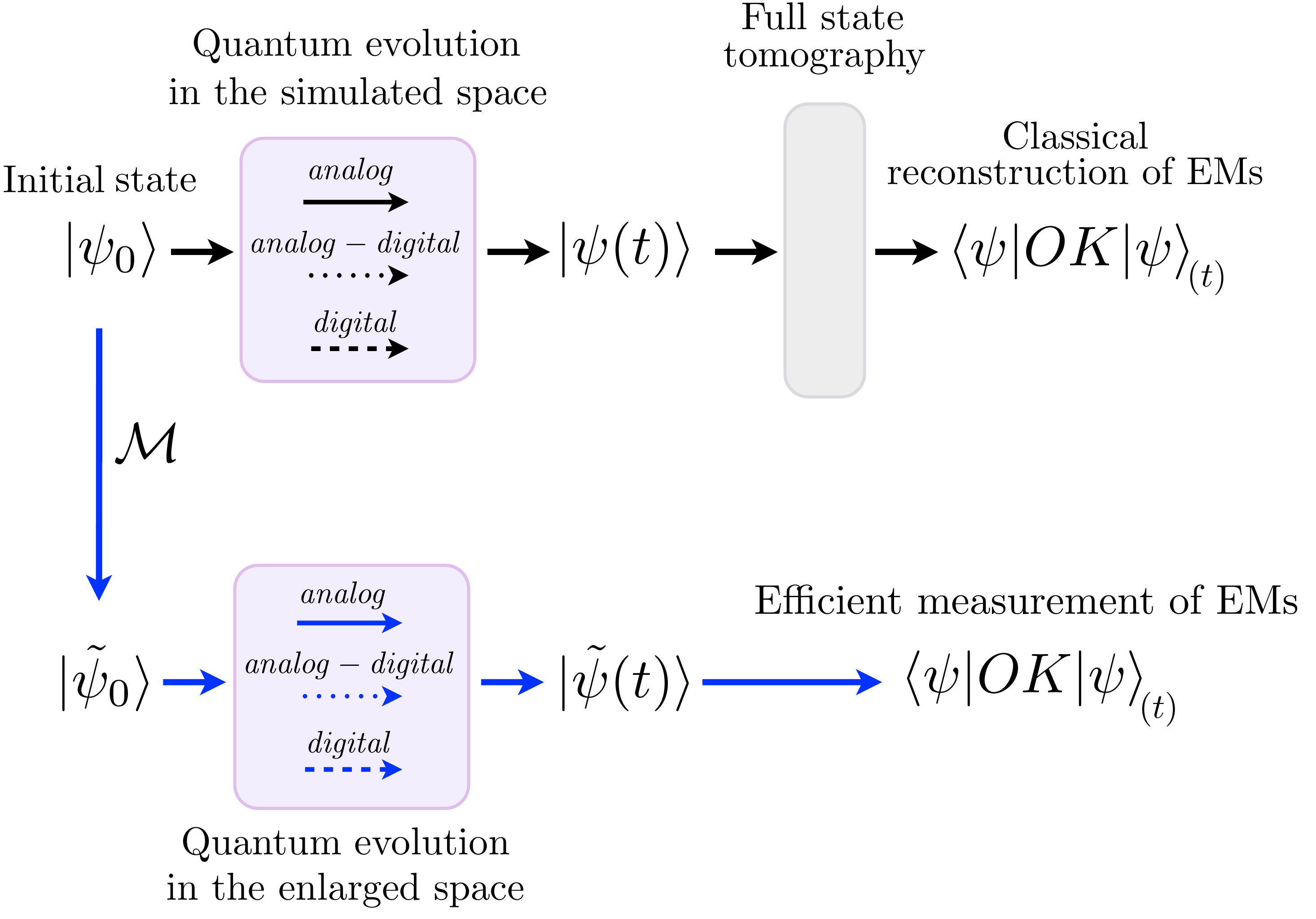}
\end{center}
\caption{(color online) Protocol for computing entanglement monotones (EMs) using the enlarged space formalism  (blue arrows), compared with the usual protocol  (black arrows). For any initial state $|\psi_0\rangle$,  we can construct throught the mapping $\mathcal{M}$ its image $|\tilde{\psi}_0\rangle$ in the enlarged space. The evolution will be implemented using analog or digital techniques giving rise to the state $|\tilde{\psi}(t)\rangle$. The subsequent  measure of a reduced number of observables  will provide us with the EMs.}\label{figscheme}
\end{figure}

So far, we have found a mapping for quantum states and expectation values between the simulated space and the simulating space. If we also want to consider an associated quantum dynamics, we would need to map the Schr\"odinger equation~(\ref{Schro}) onto another one in the enlarged space. In this sense, we look for a wave equation
\begin{equation}\label{enlarSchro}
(i\partial_t-\tilde H)|\tilde\psi(t)\rangle=0 ,
\end{equation}
whose solution  respects $|\psi(t)\rangle=M|\tilde\psi(t)\rangle$ and  $|\psi^*(t)\rangle=M\tilde{K}|\tilde\psi(t)\rangle$, thereby assuring that  the complex conjugate operation can be applied at any time $t$ with the same single qubit gate. If we define in the enlarged space a (Hermitian) Hamiltonian $\tilde H$ satisfying $M\tilde H=HM$, while applying $M$ to both sides of Eq.~\eqref{enlarSchro}, we arrive to equation $(i\partial_t-H)M|\tilde\psi(t)\rangle=0$. It follows that if $|\tilde \psi (t)\rangle$ is the solution of Eq.~\eqref{enlarSchro} with the initial condition $|\tilde\psi_0\rangle$, then $M|\tilde\psi(t)\rangle$ is the solution of the original Schr\"odinger equation~\eqref{Schro} with the initial condition $M|\tilde{\psi}_0\rangle$. Thus, if $|\psi_0\rangle=M|\tilde\psi_0\rangle$, then $|\psi(t)\rangle=M|\tilde\psi(t)\rangle$, as required. The Hamiltonian $\tilde H$ satisfying $HM=M\tilde H$ reads 
\begin{equation}\label{H}
\tilde H=\left(
\begin{array}{cc}
 i B & i A \\
 -i A & i B \\
\end{array}
\right)\equiv \big[i \mathbb{I}_2\otimes B-\sigma _y\otimes A\big],
\end{equation} 
where $H=A+iB$, with $A=A^\dag$ and $B=-B^\dag$  real matrices,  corresponds to the original Hamiltonian in the simulated space.  We note that $\tilde H$ is a Hermitian  imaginary  matrix,  e.g.   $H=\sigma_x\otimes \sigma_y+\sigma_x\otimes\sigma_z $ is mapped into $\tilde H=\mathbb{I}_2\otimes\sigma_x\otimes\sigma_y-\sigma_y\otimes\sigma_x\otimes \sigma_z$ which is Hermitian and imaginary. In this sense, $|\tilde\psi_0\rangle$ with real entries implies the same character for $|\tilde\psi(t)\rangle$, given that the Schr\"odinger equation is a first order differential equation with real coefficients. In this way, the complex-conjugate operator in the enlarged space $\tilde K=\sigma_z\otimes\mathbb{I}_{2^N}$ is the same at any time $t$.

On one hand, the implementation of the dynamics of Eq.~(\ref{enlarSchro}) in a quantum simulator will turn the computation of entanglement monotones into an efficient process, see Fig.~\ref{figscheme}. On the other hand, the evolution associated to Hamiltonian $\tilde{H}$ can be implemented efficiently in different quantum simulator platforms, as is the case of trapped ions or superconducting circuits~\cite{Lanyon11,Devoret13}. We want to point out that, in the most general case, the dynamics of a simulated system involving n-body interactions will require an embedding quantum simulator with (n+1)-body couplings. This represents, however, a small overhead of experimental resources. It is noteworthy to mention that the implementation of many-body spin interactions have already been realized experimentally in digital quantum simulators in trapped ions~\cite{Lanyon11}. Concluding, quantum simulations in the enlarged space require the quantum control of {\it only one additional qubit}.

{\it Efficient computation of entanglement monotones.---} A general entanglement monotone constructed with $K$, $\sigma_y$, and $g^{\mu\nu}\sigma_\mu\sigma_\nu$, contains at most $3^k$ terms of the form $\langle\psi| O K|\psi\rangle$, $k$ being the number of times that $g^{\mu\nu}\sigma_\mu\sigma_\nu$ appears. Thus, to evaluate the most general set of entanglement monotones, we  need to measure $2\cdot 3^k$ observables, in contrast with the $2^{2N}-1$ required for full tomography.

We present now examples showing how our protocol minimizes the required experimental resources. 

{\it i)} {\it The concurrence.---} This two-qubit entanglement monotone defined in Eq.~(\ref{Concurrence})  is built using  $\sigma_y$ and $K$, and it can be evaluated with the measurement of  $2$ observables  in the enlarged space, instead of the $15$ required for  full tomography. Suppose we know $|\psi_0\rangle$ and want to compute $C(|\psi(t)\rangle)$, where $|\psi(t)\rangle \equiv e^{-iHt}|\psi_0\rangle$. We first initialize the quantum simulator with the state $|\tilde\psi_0\rangle$ using the mapping of Eq.~(\ref{map}). Second, this state evolves according to Eq.~(\ref{enlarSchro}) for a time $t$. Finally, following Eq.~\eqref{obs} with $O=\sigma_y\otimes\sigma_y$, we compute the quantity 
\begin{equation}\label{defen}
\langle\tilde \psi(t)|\sigma_z\otimes\sigma_y\otimes\sigma_y-i\sigma_x\otimes\sigma_y\otimes\sigma_y|\tilde\psi(t)\rangle ,
\end{equation}
by measuring the  observables $\sigma_z\otimes\sigma_y\otimes\sigma_y$ and $\sigma_x\otimes\sigma_y\otimes\sigma_y$ in the enlarged space.

{\it ii) The $3$-tangle.---} The $3$-tangle~\cite{Dur00} is a $3$-qubit entanglement monotone defined as $\tau_3(|\psi\rangle)=|g^{\mu\nu}\langle\psi|\sigma_\mu\otimes\sigma_y\otimes\sigma_yK|\psi\rangle\langle\psi|\sigma_\nu\otimes\sigma_y\otimes\sigma_yK|\psi\rangle|$. It is built using  $g^{\mu\nu}\sigma_\mu\sigma_\nu$ and $K$, so the computation of $\tau_3$ in the enlarged space requires $6$ measurements  instead of the $63$ needed for full-tomography. The evaluation of $\tau_3(|\psi(t)\rangle)$ can be achieved  following  the same steps explained in the previous example, but now computing the quantity
\begin{eqnarray}
\big|&&-\langle\tilde\psi(t)|\sigma_z\otimes\mathbb{I}_2\otimes\sigma_y\otimes\sigma_y-i\sigma_x\otimes\mathbb{I}_2\otimes\sigma_y\otimes\sigma_y|\tilde\psi(t)\rangle^2  \nonumber \\ 
&&+\langle\tilde\psi(t)|\sigma_z\otimes\sigma_x\otimes\sigma_y\otimes\sigma_y-i\sigma_x\otimes\sigma_x\otimes\sigma_y\otimes\sigma_y|\tilde\psi(t)\rangle^2  \nonumber \\
&&+\langle\tilde\psi(t)|\sigma_z\otimes\sigma_z\otimes\sigma_y\otimes\sigma_y-i\sigma_x\otimes\sigma_z\otimes\sigma_y\otimes\sigma_y|\tilde\psi(t)\rangle^2\big| , \nonumber \\
\end{eqnarray}
with  the corresponding measurement of  observables in the enlarged space.

{\it iii) N-qubit monotones.---}
In this case, the simplest entanglement monotone is $|\langle\psi|\sigma_y^{\otimes N}K|\psi\rangle|$ if $N$ is even (expression that is identically zero if $N$ is odd), and $|g^{\mu\nu}\langle\psi|\sigma_\mu\otimes\sigma_y^{\otimes N-1}K|\psi\rangle\langle\psi|\sigma_\nu\otimes\sigma_y^{\otimes N-1}K|\psi\rangle|$ if $N$ is odd. The first entanglement monotone needs $2$ measurements, while the second one needs $6$. This minimal requirements have to be compared with the $2^{2N}-1$ observables  required for full quantum tomography.

{\it iv) The mixed-state case.---}
Once we have defined  $E(|\psi\rangle)$ for the pure state case, we can extend our method to the mixed state case via the convex roof construction, see Eq. (\ref{roof}). Such a definition  is needed  because the possible  pure state decompositions of $\rho$ are infinite, and each of them brings a different value of $\sum_{i}p_iE(|\psi_i\rangle)$.  By considering its minimal value, as in Eq.~(\ref{roof}), we eliminate this ambiguity preserving the properties that define an entanglement monotone. To decide when $E(\rho)$ is zero  is called {\it separability problem}, and it is proven to be NP-hard for states close enough  to the border between the sets of entangled and separable states~\cite{Gurvits1, Gharibian1}. However, there exist useful classical algorithms~\cite{footnote} able to find an estimation of $E(\rho)$ up to a finite error~\cite{Loss1, Loss2}.

Our approach for mixed states involves a hybrid quantum-classical algorithm, working well in cases in which $\rho$ is approximately a low-rank state. We restrict our study to the case of unitary evolutions acting on mixed-states, given that the inclusion of dissipative processes would require an independent development. Let us consider a state with rank $r$ and assume that the pure state decomposition solving Eq. \eqref{roof} has $c$ additional terms. That is, $k=r+c$, with $k$ being the number of terms in the optimal decomposition, while $c$ is assumed to be low. An algorithm that solves Eq.~\eqref{roof} (see for example~\cite{Loss1, Loss2}) evaluates at each step the quantity $\sum_{i=1}^kp_iE(|\psi_i\rangle)$ and, depending on the result, it changes $\{p_i,|\psi_i\rangle\}$ in order to find the minimum. Our method consists in inserting an embedded quantum simulation protocol in the evaluation of each $E(|\psi_i\rangle)$, which can be done with few measurements in the enlarged space. We gain in efficiency with respect to full tomography if  $k\cdot l\cdot m<2^{2N}-1$, where $l$ is the number of iterations of the algorithm and $m$ is the number of measurements to evaluate the specific entanglement monotone. We note that $m$ is a constant that can be low, depending on the choice of $E$, and, if $\rho$ is low rank, $k$ is a low constant too. With this approach, the performance of the computation of entanglement monotones, $E ( \rho)$, can be cast in two parts: while the quantum computation of $\sum_{i=1}^kp_iE(|\psi_i\rangle)$ can be efficiently implemented, the subsequent minimization remains a difficult~task.

{\it Conclusions.---} We have presented a paradigm for the efficient computation of a class of entanglement monotones requiring minimal experimental added resources. The proposed framework consists in the adequate embedding of  a quantum dynamics in the degrees of freedom of an enlarged-space quantum simulator.  In this manner, we have proposed novel concepts merging the fundamentals of quantum computation with those of quantum simulation. We believe that this novel embedding framework for quantum simulators will enhance the capabilities of one-to-one quantum simulations.

The authors acknowledge support from Spanish MINECO FIS2012-36673-C03-02; UPV/EHU UFI 11/55; UPV/EHU PhD fellowship; Basque Government IT472-10; SOLID, CCQED, PROMISCE, SCALEQIT EU projects; and Marco Polo PUCP grant.


\begin{thebibliography}{10}
\bibitem{footnote} We stress that this is not equivalent to solving the separability problem, as the classical algorithm will always give a strict upper bound of $E(\rho)$.
\end{thebibliography}
\end{document}